\documentclass[%
 reprint,
 amsmath,amssymb,
 aps,
]{revtex4-2}

\usepackage[normalem]{ulem}
\usepackage{xcolor}
\usepackage{graphicx}
\usepackage{dcolumn}
\usepackage{bm}
\usepackage[utf8]{inputenc}

\begin{document}

\preprint{APS/123-QED}

\title{Nontrivial Twisted States in Nonlocally Coupled Stuart-Landau Oscillators}

\author{Seungjae Lee}
 \email{seungjae.lee@tum.de}
\affiliation{Physik-Department, Technische Universit\"at M\"unchen, James-Franck-Stra\ss e 1, 85748 Garching, Germany}

\author{Katharina Krischer}%
 \email{krischer@tum.de}
\affiliation{Physik-Department, Technische Universit\"at M\"unchen, James-Franck-Stra\ss e 1, 85748 Garching, Germany}%

\date{\today}

\begin{abstract}

A twisted state is an important yet simple form of collective dynamics in an oscillatory medium. Here, we describe a nontrivial type of twisted state in a system of nonlocally coupled Stuart-Landau oscillators. The nontrivial twisted state (NTS) is a coherent traveling wave characterized by inhomogeneous profiles of amplitudes and phase gradients, which can be assigned a winding number. To further investigate its properties, several methods are employed. We perform a linear stability analysis in the continuum limit and compare the results with Lyapunov exponents obtained in a finite-size system. The determination of covariant Lyapunov vectors allows us to identify collective modes. Furthermore, we show that the NTS is robust to small heterogeneities in the natural frequencies and present a bifurcation analysis revealing that NTSs are born/annihilated in a saddle-node bifurcation and change their stability in Hopf bifurcations. We observe stable NTSs with winding number 1 and 2. The latter can lose stability in a supercritical Hopf bifurcation, leading to a modulated 2-NTS.

\end{abstract}

\maketitle


\section{\label{sec:intro}INTRODUCTION}

The collective dynamics of an ensemble of coupled oscillators is key to the functioning of many systems in practically all scientific fields~\cite{pikovksy_sync,strogatz_sync}.
Accordingly, the dynamics of ensembles of oscillators with different couplings has been studied intensively during the last decades. One coupling topology that has proven important for discovering new synchronization patterns and revealing the mechanisms that give rise to them is nonlocal coupling in a ring geometry, the most prominent example being a chimera state~\cite{kuramoto2002,abrams2004}. When the coupling between the oscillators is weak, the dynamics can be captured by considering only the evolution equations of the phases of the oscillators~\cite{Omel_chenko_2013,Panaggio_2015,Omel_chenko_2018}. For somewhat stronger coupling, the amplitudes of a part of the oscillators along the ring might exhibit variations decisive for the dynamics, as found in amplitude-mediated chimera states~\cite{amc,amc2,amc3,amp-chimera}.

Another prominent and compared to the chimera state simpler collective dynamics found in a ring of nonlocally coupled oscillators is a so-called twisted state~\cite{pikovsky-twisted,omelchenko-twisted,sjlee_twisted,bains_splay,maistrenko_twisted}. In a `traditional' twisted state, the phase difference between adjacent oscillators is always the same such that the phase winds around the ring an integer multiple of $2\pi$ whereas the amplitude of all the oscillators attains the same constant value. A twisted state has thus been seen as a typical phenomenon that is fully captured by a phase-reduced model. The phase profile evolves according to $\phi(x,t) = \frac{2\pi q}{L}x +\Omega t$, where $\Omega$ is a collective frequency and $L$ is the length of the medium. Correspondingly, the phase gradient is everywhere given by $\partial_x \phi(x,t) = \frac{2\pi q}{L}= \textrm{const.}$ with $q \in \mathbb{Z}$ defining a winding number, while the trivial amplitude dynamics obeys $r(x,t) = r_0 \in \mathbb{R}_{>0}$ for all $x \in [0,L]$. 

In this paper, we show that in a ring of nonlocally coupled oscillators another type of a twisted state might form. This state is characterized by a non-constant gradient of the phase profile and an inhomogeneous amplitude profile which travels along the ring with a fixed shape and a constant speed. The phase still advances by a multiple of $2\pi$ when going once around the ring so that the solution can be characterized by a winding number and the state be considered a twisted state. 
Due to the spatio-temporal variations of the amplitude the state does not exist in the classical phase-reduced model but its description requires a priori planar oscillators. In order to contrast this novel type of twisted state from the so far known constant phase-gradient and constant amplitude twisted state, we coin a twisted state with non-uniform amplitude and phase gradient profiles a \textit{nontrivial twisted state} (NTS) and a twisted state with uniform profiles of amplitude and phase gradient a \textit{trivial twisted state} (TTS).  Both these states are discussed in the following sections with a system of nonlocally coupled identical Stuart-Landau oscillators in a 1-d ring, with an emphasis on the dynamical and spectral properties of the NTS.

In Sec.~\ref{sec:traveling wave}, we first discuss the dynamical properties of NTSs in the original space and time coordinates. Then we perform a linear stability analysis in a moving and co-rotating reference frame where both amplitude and phase profiles become stationary. Finally, we compare the stability results with those of the TTS. In Sec.~\ref{sec:lyapunov}, we address the stability of finite-size ensembles and study how the spectral properties of the finite-size state converge to those of the continuum limit as the system size increases. The stability of the finite-size ensemble is obtained from the numerical determination of the \textit{Lyapunov exponents} (LEs). In addition, \textit{covariant Lyapunov vectors} (CLVs) are considered to confirm the existence of collective Lyapunov modes~\cite{pikovsky_LE,LE1,sjlee1,clv1,clv2,clv3}. 
Next, in Sec.~\ref{sec:hetero}, we demonstrate that the NTS is robust with respect to small heterogeneities in the natural frequencies of the Stuart-Landau oscillators~\cite{pikovsky-chimera2021,omelchenko-twisted}. Finally, a bifurcation analysis is performed in Sec.~\ref{sec:bifurcation}, following the procedure described in~\cite{Laing-continuation}. It reveals that the NTS is stable in a large parameter range. We summarize the results in Sec.~\ref{sec:conclusion}.

\section{\label{sec:traveling wave}Nontrivial Twisted State}

\subsection{\label{subsec:govern-eqn} Governing Equation and Observable Dynamics}

We consider nonlocally coupled, identical Stuart-Landau oscillators along a 1-d ring of length $L$. The oscillators are described by complex-valued dynamical agents $W(x,t) =r(x,t)e^{i\phi(x,t)} \in \mathbb{C}$ where $x \in [0,L]$. The oscillator field is governed by
\begin{flalign}
\frac{\partial}{\partial t}W(x,t) &= \mathcal{F}(W(x,t)) +\varepsilon e^{-i \alpha(H(x,t))} H(x,t) \notag \\
&=  (1+i\omega)W(x,t) -|W(x,t)|^2 W(x,t) \notag \\
&+\varepsilon e^{-i \alpha(H(x,t))} \int_0^L G(x-x') W(x',t) dx'
\label{Eq:complex-governing}
\end{flalign} where periodic boundary conditions are imposed. The uncoupled local dynamics is given by a Stuart-Landau oscillator $\mathcal{F}(W)=(1+i\omega)W -|W|^2 W$ and the nonlinear phase-lag function is assumed to be $\alpha(H(x,t)) = \alpha_0 + \alpha_1 | H(x,t) |^2$ with real parameters $\alpha_0$, $\alpha_1 \in \mathbb{R}$ ~\cite{pikovsky-chimera2018,pikovsky-breathing2017,pikovsky-SL2010}. The coupling strength $\varepsilon$ is a real parameter and the frequency of the identical oscillators is set to $\omega=0$. 

The forcing field is defined as an integral convolution operator, i.e.,
\begin{flalign}
H(x,t) = (\mathcal{G}W)(x,t) := \int_0^L G(x-x') W(x',t) dx'.
\label{Eq:forcing-field}
\end{flalign} The nonlocal coupling kernel given by 
\begin{equation}
    G(y) = \frac{\kappa}{2 \textrm{sinh}(\kappa L/2)} \textrm{cosh} \bigg( \kappa (|y|- L/2) \bigg) 
    \label{Eq:Coupling-kernel}
\end{equation} for $|y| \leq L/2$ so that both the normalization condition $\int_{-L/2}^{L/2} G(y) dy = 1$ and the Green's function of the inhomogeneous Helmholtz equation~\cite{pikovsky-chimera2018,pikovsky_2017}
\begin{equation}
    (\partial_{x}^{2} -\kappa^2) H(x,t)=-\kappa^2 W(x,t)
    \notag
\end{equation}
are satisfied with the periodic boundary conditions: $H(0,t)=H(L,t)$ and $\partial_x H(0,t) = \partial_x H(L,t)$. Note that $\kappa$ is a real parameter and $\kappa^{-1}$ determines the coupling range and has the dimension of a length. Thus, $\kappa^{-1}$ also characterizes the length of the medium~\cite{pikovsky_2017,pikovsky-chimera2018,pikovsky-chimera2021}. In the limit of $\kappa L \rightarrow \infty$, the coupling kernel becomes $G(x)=\kappa e^{-\kappa |x| }/2$, as used in \cite{kuramoto2002}. In the following (up to Sec.~\ref{sec:bifurcation}), we use the following parameter values: $\varepsilon = 1$, $\kappa=4.874$, $L=1$, $\alpha_0 = -0.4 \pi$ and $\alpha_1 = -(\pi/2+\alpha_0)/0.36$. Note that $\varepsilon$ is no longer small such that the amplitude variables may follow nontrivial dynamics.

\begin{figure}[t!]
\includegraphics[width=1.0\linewidth]{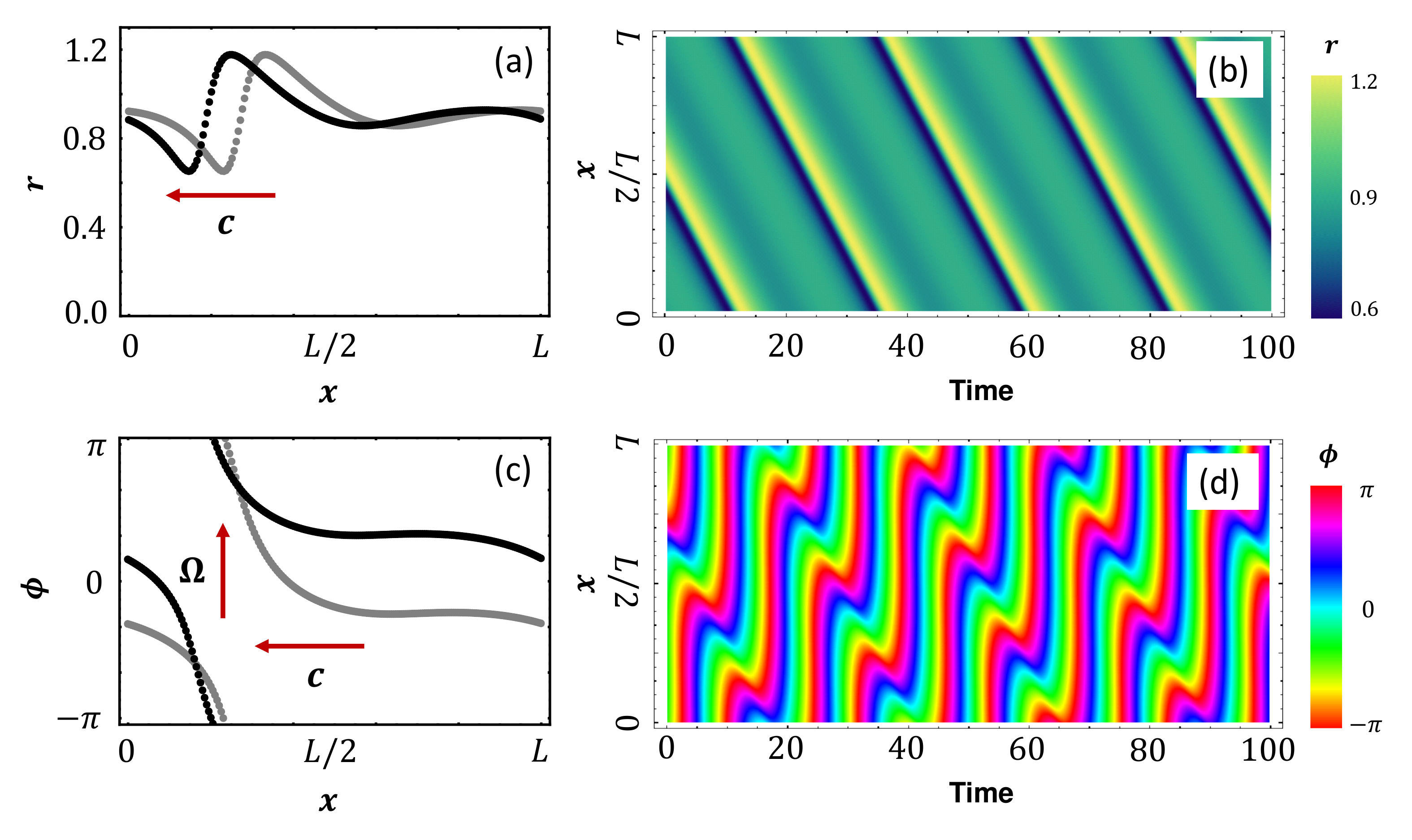}
\caption{\label{Fig:TW-snapshot} Nontrivial twisted state dynamics obtained from a random initial condition. (a) Amplitude profiles at $t= 10^4$ (gray) and $t+2$ (black), (b) spatio-temporal evolution of the amplitude ($t \geq 10^5$), (c) phase profiles at $t= 10^4$ (gray) and $t+2$ (black), (d) spatio-temporal evolution of the phase ($t \geq 10^5$). Other parameter values: $\varepsilon = 1$, $\kappa=4.874$, $L=1$, $\alpha_0 = -0.4 \pi$ and $\alpha_1 = -(\pi/2+\alpha_0)/0.36$.}
\end{figure}

\begin{figure*}[t!]
\includegraphics[width=1.0\textwidth]{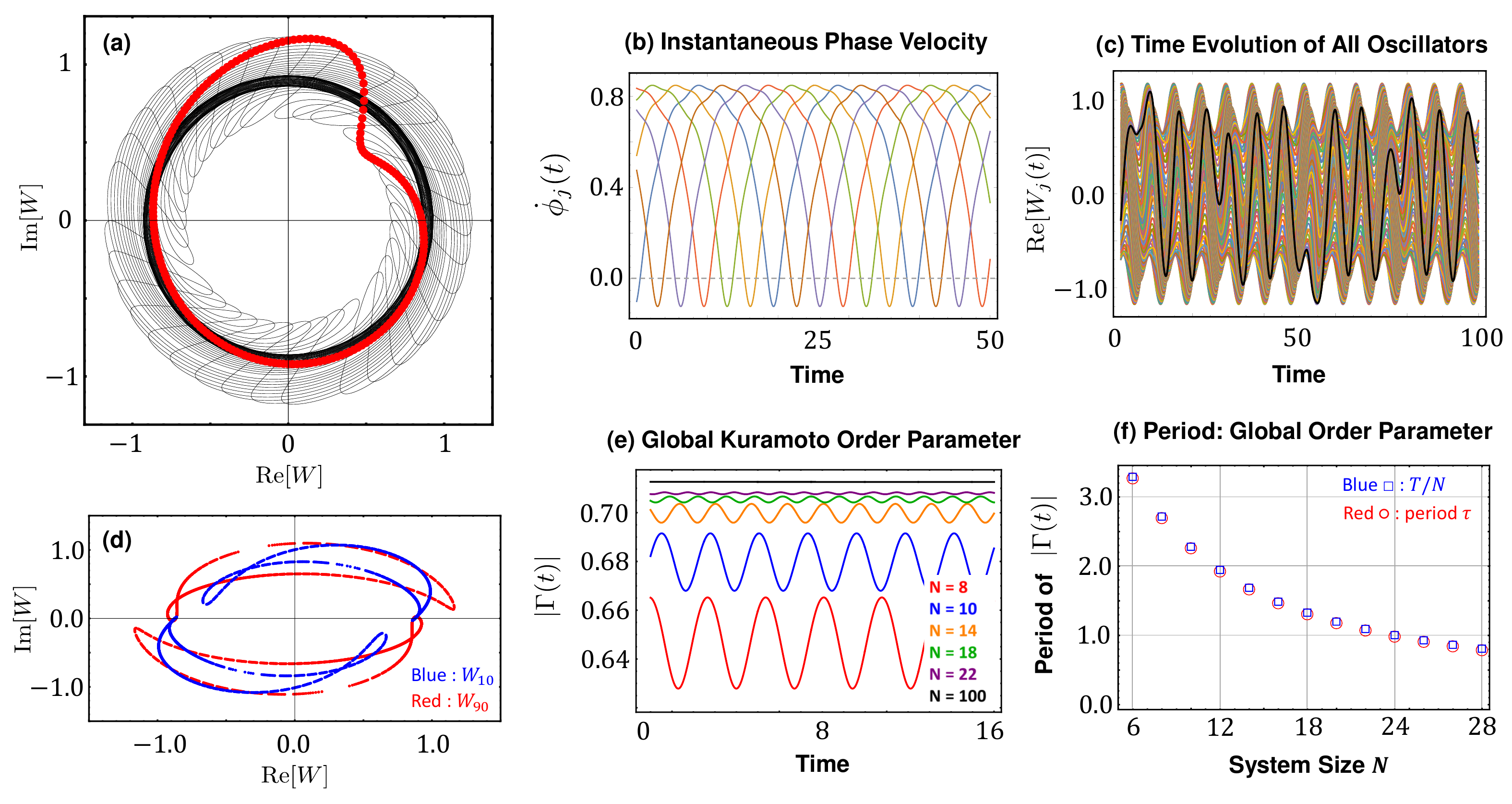}
\caption{ (a) Red curve: Snapshot of the amplitude profile of an NTS on the ring, i.e., $r(\frac{2\pi}{L} x)\textrm{cos}(\frac{2\pi}{L} x)$ vs. $r(\frac{2\pi}{L} x)\textrm{sin}(\frac{2\pi}{L} x)$ for $x \in [0,L]$ and $N=200$; black line: trajectory of the first oscillator in the complex plane: $\textrm{Re}[W_1(t)]$ vs. $\textrm{Im}[W_1(t)]$); (b) Instantaneous phase velocity as a function of time for $N=6$. (c) Time evolution of $\textrm{Re}[W(t)]$ for $N=200$. The black line highlights the time series of one of the oscillators.  (d) Poincar\'e map of $W_{10}$ and $W_{90}$ in the complex plane where the Poincar\'e section is defined by $\phi_1(t) \equiv 0$ $(\textrm{mod}2\pi)$. (e) Modulus of the global Kuramoto order parameter as a function of time for different system sizes $N$. All numerical values shown for $t \geq 10^4$. (f) Period of the modulus of the global Kuramoto order parameter as a function of the system size $N$. The numerically obtained value (red) coincides with $T/N$ (blue).} 
\label{Fig:TW-dynamics}
\end{figure*}

For the chosen parameter values, the microscopic dynamics of the finite-size approximation with $N = 200$ oscillators may exhibit an NTS along the ring, as depicted in Fig.~\ref{Fig:TW-snapshot}. In (a,b) the amplitude dynamics $r_j(t)$ and in (c,d) the phase dynamics $\phi_j(t)$ is shown where $r_j(t)e^{i \phi_j(t)} = W_j(t)$ at $x_j = \frac{j-1}{N-1} \in [0,1]$ such that $\phi_j(t)$ and $r_j(t)$ are governed by
\begin{flalign}
  \frac{d \phi_j}{dt} 
  &= \omega + \frac{\varepsilon}{r_j} \textrm{Im} \left[ H_j(t) e^{-i \phi_j} e^{-i\alpha(H_j(t))} \right] 
  \label{Eq:phase-governing-SLE}
\end{flalign} and
\begin{flalign}
  \frac{d r_j}{dt} 
  &= r_j-r_j^3 + \varepsilon \textrm{Re} \left[ H_j(t) e^{-i \phi_j} e^{-i\alpha(H_j(t))} \right] 
\label{Eq:amp-governing-SLE}
\end{flalign} and $H_j(t) = H(x_j,t)$ for $j=1,2,...,N$. 

As apparent from the two amplitude snapshots in Fig.~\ref{Fig:TW-snapshot}(a), the amplitudes form a smooth time-dependent curve as a function of $x$. The spatio-temporal evolution of the amplitude profiles shown in Fig~\ref{Fig:TW-snapshot}(b) evidence that the profiles travel along the ring with a fixed shape and a constant speed. The amplitude dynamics thus constitutes a traveling wave solution. As apparent from the two snapshots of the phase profiles depicted in Fig.~\ref{Fig:TW-snapshot} (c), the phase appears to be smooth along $x$ and exhibits large variations in the region where the amplitude variations are large, and shallow variations where the amplitude varies only slightly. If we define the phase difference  modulo $2\pi$ in the interval $ [-\pi,\pi)$ as $\Delta_{i,j}:=\phi_{i}-\phi_j$ with $\phi_{N+1}\equiv \phi_1$, then we can assign a winding number to the observed state according to $q=\frac{1}{2\pi}\sum_{j=1}^{N}\Delta_{j+1,j}\in \mathbb{Z}$. In the example shown in Fig.~\ref{Fig:TW-snapshot}, $q=-1$ (but note that depending on the initial condition $q$ can also be $+1$). Besides, $\Delta_{j+1,j} \neq \textrm{const}$. The dynamics thus constitutes an NTS as defined in the introduction. Furthermore, the phase profile is uniformly rotating with a collective frequency $\Omega$ and, like the amplitude profile, it travels to the left with the lateral speed $c$ (Fig.~\ref{Fig:TW-snapshot} (c,d)). In an appropriately rotating frame, an NTS thus constitutes a traveling wave, just as a TTS does, which is a coexisting solution at the same parameter values, as demonstrated below.

In order to validate that the NTS is in fact a coherent traveling wave, as a TTS is, we determined the Kuramoto local order parameter $z(x,t)$ defined as~\cite{chimera-spectral,Omelchenko2022}
\begin{equation}
    z(x,t) =\lim_{N \rightarrow \infty} \frac{1}{|B_{\delta}^{N}(x)|} \sum_{j \in B_{\delta}^{N}(x) } e^{i \phi_j(t)}  
    \label{Eq: local order paramter}
\end{equation} with $B_{\delta}^{N}(x) = \{j: 1 \leq j \leq N, |x-x_j| <\delta \}$ for small enough $0<\delta \ll 1$. Equation (\ref{Eq: local order paramter}) directly shows how to calculate the local order parameter numerically from the finite-size microscopic dynamics. In the continuum limit, Equation (\ref{Eq: local order paramter}) is equivalent to the more intuitive version defined for a spatially extended 1-dimensional system, which reads
\begin{flalign}
  z(x,t) = \frac{1}{2\delta} \int_{x-\delta}^{x+\delta} e^{i\phi(x',t)} dx'
\end{flalign} for $0 < \delta \ll 1$~\cite{pikovsky-twisted,pikovsky-chimera2018}. Hence, the local order parameter provides a coarse-grained macroscopic observable that is continuous both in $x$ and $t$, even though $\phi(x,t)$ in general is not, and characterizes a local degree of coherence in a small neighborhood around $x$~\cite{Omel_chenko_2013,Omel_chenko_2018,pikovsky-WS1}. In fact, we obtain $ |z(x,t)|=1$ for all $x \in [0,1]$, ensuring that the NTS is a coherent traveling wave.

In Fig.~\ref{Fig:TW-dynamics} (a), the trajectory of one of the oscillators (black line) is depicted together with a snapshot of the amplitude profile (red curve) in the complex plane. The trajectories encircle the origin, but exhibit a backward motion in phase when the amplitude of the oscillation exhibits a pronounced deformation from a circular structure (akin to the \textit{apparent retrograde motion} of a planet from the earth's viewpoint) (cf. also Fig.~\ref{Fig:TW-snapshot} (d)). The reversal of the direction of phase change reflects the negative values the instantaneous phase velocity attains when the oscillation amplitude goes through the hump (Fig.~\ref{Fig:TW-dynamics} (b)). The time evolution of $\textrm{Re}(W)$ of the oscillators is shown in Fig.~\ref{Fig:TW-dynamics} (c). We notice that each individual oscillator exhibits some apparently irregular oscillation (illustrated by the black highlighted curve)  while the motion of the entire ensemble displays a periodically oscillating envelope. In Fig.~\ref{Fig:TW-dynamics} (d), the trajectories of two representative oscillators ($W_{10}$ and $W_{90}$) are depicted in a Poincar\'e section defined by $\phi_1(t) \equiv 0$ (mod~$2\pi$). Clearly, all points of the trajectory of each oscillator lie on two closed curves which reveals that the oscillators exhibit in fact a quasi-periodic motion in phase space.

Further dynamical properties of the NTS can be derived from the modulus of the \textit{global Kuramoto order parameter} $\Gamma$ defined as
\begin{equation}
\Gamma (t) :=  \frac{1}{L} \int_{0}^{L} e^{i\phi(x,t)}dx
\label{Eq:global-Kuramoto-OP}
\end{equation} which corresponds to $\frac{1}{N}\sum_{j=1}^{N}e^{i \phi_j(t)}$ in the finite-size approximation. In Fig.~\ref{Fig:TW-dynamics} (e), $|\Gamma|$ exhibits a periodic motion for small system size $N$ whereby the period and amplitude of its oscillations decrease as $N$ increases. For large $N$, $|\Gamma|$ eventually attains a constant value, and $\Gamma(t)$ rotates uniformly with $\Omega$, i.e., $\Gamma(t)=|\Gamma| e^{i \Omega t}$. The behavior of $|\Gamma|$ with system size can be understood from the observation that the instantaneous phase velocities $\{\dot{\phi}_{i}(t)\}_{i=1}^{N}$ of all oscillators are periodic functions with the same period $T \approx 23$, and identical shapes while being shifted in time by equal amounts, as can be seen in Fig.~\ref{Fig:TW-dynamics} (b). A similar phenomenon was reported for a so-called Poisson chimera state in a two-population network~\cite{sjlee1}. Following the same argument as given therein, it is assumed that $\dot{\phi}_{i}(t-\frac{j}{N}T)=\dot{\phi}_{i+j}(t)$ for an arbitrary $j \in \{1,...,N\}$, which gives $\phi_{i}(t-\frac{T}{N}) = \phi_{i+1}(t)+\Theta_0$ for $i=1,...,N$ with a common constant shift $\Theta_0 \in \mathbb{R}$ and $\phi_{1} \equiv \phi_{N+1}$. Substituting this into Eq.~(\ref{Eq:global-Kuramoto-OP}),
\begin{flalign}
  |\Gamma(t)| &= \bigg|\frac{1}{N}\sum_{j=1}^{N}e^{i\phi_{j+1}(t)} \bigg| = \bigg|\frac{e^{- i\Theta_0}}{N}\sum_{j=1}^{N}e^{i\phi_{j}(t-\frac{T}{N})}  \bigg| \notag \\ &= |\Gamma(t-\frac{T}{N})| = |\Gamma(t-\tau)| \notag
\end{flalign}
we obtain $|\Gamma(t)| = |\Gamma(t-\frac{T}{N})|$ for $\forall_t$, that is, the modulus of the global order parameter is periodic with the period $\tau = \frac{T}{N}$ decreasing with increasing $N$. This is also numerically verified in Fig.~\ref{Fig:TW-dynamics} (f): The period of $|\Gamma|$ numerically obtained from Eq.~(\ref{Eq:global-Kuramoto-OP}) coincides with $T/N$, i.e., the period $T$ of each instantaneous phase velocity divided by the system size $N$. Hence, the modulus of the global Kuramoto order parameter of an NTS has a non-zero constant value for a sufficiently large system size and oscillates around a non-zero mean for small system sizes. This is in contrast to a TTS which has a zero global Kuramoto order parameter $|\Gamma(t)| = 0$ for all $N \in \mathbb{N}$.

\subsection{\label{subsec:stability}Linear Stability of the Nontrivial Twisted States}

The linear stability of an NTS can be obtained by going to a reference frame moving with a constant speed $c$ and rotating uniformly with $\Omega$. In this reference frame, both phase and amplitude profiles are stationary. Therefore, we make the ansatz
\begin{flalign}
W(x,t)=W_0(\xi)e^{i\Omega t}
\end{flalign} where $\xi=x-ct$. The winding number of the NTS is then given by $q = \frac{1}{2\pi} \sum_{j=1}^{N}(\Phi_{j+1}-\Phi_j)$ where $\Phi_j = \textrm{arg}[W_0(\xi_j)]$ at $\xi_j=\frac{(j-1)}{N-1} \in [0,1]$ for $j=1,...,N$, and the NTS satisfies
\begin{flalign}
-c \partial_\xi W_0(\xi) &= (1+ i \Delta) W_0(\xi) -|W_0(\xi)|^2W_0(\xi) \notag\\
&+\varepsilon e^{-i \alpha((\mathcal{G}W_0)(\xi))}(\mathcal{G}W_0)(\xi) \label{Eq:complex-stationary-profile}
\end{flalign} where $\partial_\xi := \frac{d}{d\xi}$ and $\Delta = \omega-\Omega$ is a real unknown constant. Here, the integral convolution operator reads
\begin{flalign}
(\mathcal{G}W_0)(\xi) = H_0(\xi) = \int_{0}^{L}G(\xi-\xi')W_0(\xi')d\xi' \label{Eq:complex-forcing-field-moving}
\end{flalign} where $G(y)$ is defined in Eq.~(\ref{Eq:Coupling-kernel}). 

Since in the reference frame defined above the NTS is a stationary solution, we can obtain its linear stability by linearizing the evolution equation around the stationary wave profile and determining the eigenvalues of the linearized equations. To do so, we first consider the coordinate transformation
\begin{equation}
    W_0(\xi) = X_0(\xi) +i Y_0(\xi) \notag
\end{equation} where $\textrm{Re}W_0=X_0$ and $\textrm{Im}W_0=Y_0$ are real-valued functions that are periodic in $\xi$: $X_0(\xi+L)=X_0(\xi)$ and $Y_0(\xi+L)=Y_0(\xi)$. Then, we rewrite  Eq.~(\ref{Eq:complex-stationary-profile}) as follows
\begin{flalign}
 -c \partial_\xi \begin{pmatrix}
 X_0\\
 Y_0
\end{pmatrix} 
&= \Bigg[
\begin{pmatrix}
1 & -\Delta\\
\Delta & 1
\end{pmatrix} -(X_0^2+Y_0^2) I_2 \Bigg] 
\begin{pmatrix}
 X_0\\
 Y_0
\end{pmatrix} \notag \\
&+\varepsilon \begin{pmatrix}
\textrm{cos}\alpha & \textrm{sin}\alpha\\
-\textrm{sin}\alpha & \textrm{cos}\alpha
\end{pmatrix}\begin{pmatrix}
 (\mathcal{G}X_0)(\xi)\\
 (\mathcal{G}Y_0)(\xi)
\end{pmatrix} , \notag  \\ \notag \\\alpha &= \alpha_0 + \alpha_1 |H_0(\xi)|^2
\label{Eq:real-valued-moving-reference}
\end{flalign} where $I_2$ is a $2 \times 2$ identity matrix. Next, we consider a small deviation from $W_0(\xi)$: $v_1=X(\xi,t)-X_0(\xi)$ and $v_2=Y(\xi,t)-Y_0(\xi)$ with $|v_i| \ll 1$ for $i=1,2$. Note that we treat $\xi$ here as a time-independent spatial variable. Then, the linearized equation is given by
\begin{equation}
    \frac{dV}{dt} = \mathcal{L}V
    \label{Eq: linear operator equation}
\end{equation} where $V = (v_1,v_2)^\top$ and $\mathcal{L}:=\mathcal{M}+\mathcal{K}$ is a time-independent linear operator that governs the tangent space dynamics of the perturbation whose point and continuous spectra $\sigma(\mathcal{L}) = \sigma_{\textrm{pt}}(\mathcal{L}) \cup \sigma_{\textrm{cont}}(\mathcal{L})$ determine the linear stability of the profiles of the NTS. To numerically investigate the spectral properties of an NTS profile, we consider uniformly discretized operators that are calculated at each $\xi=\xi_j$ for $j=1,...,M$ with $M \gg 1$~\cite{pikovsky_2017}. The operator given by a multiplication then reads
\begin{flalign}
 (\mathcal{M}V)(\xi) &= \mathcal{M}(\xi)V(\xi) =\Bigg[ \begin{pmatrix}
c \mathcal{D}+2Y_0^2 & -\Delta\\
\Delta & c \mathcal{D}+2X_0^2
\end{pmatrix} \notag \\
&+\begin{pmatrix}
\textrm{Re}~\eta(\xi)  & \textrm{Im}~\eta(\xi)\\
\textrm{Im}~\eta(\xi) &\textrm{Re}~\eta(\xi)  
\end{pmatrix} \Bigg] V(\xi), \notag  \\ \notag \\
\eta(\xi) &= 1-3(X_0^2+Y_0^2)-i2X_0 Y_0 \notag
\end{flalign} where $\mathcal{D} \equiv \partial_\xi$ is a differential operator which in our numerical approach we evaluate spectrally following Ref.~\cite{spectral}. It is approximately treated as a constant matrix operator. From the numerical evaluation of the eigenvalues of $\mathcal{M}$ it follows that we obtain only a discretization of continuous eigenvalue branches so that $\sigma(\mathcal{M})=\sigma_{\textrm{cont}}(\mathcal{M})$ holds~\cite{Omel_chenko_2013}. On the other hand, the compact integral operator $\mathcal{K}$ is given by
\begin{flalign}
   (\mathcal{K}V)(\xi) &= \varepsilon \bigg( \mathcal{A}(\xi)+2\alpha_1 \mathcal{B}(\xi)\bigg) \begin{pmatrix}
 (\mathcal{G}v_1)(\xi)\\
 (\mathcal{G}v_2)(\xi)
\end{pmatrix}, \notag \\ \notag \\
\mathcal{A}(\xi) &= \begin{pmatrix}
\textrm{cos}\alpha  & \textrm{sin}\alpha \\
-\textrm{sin}\alpha  & \textrm{cos}\alpha  \end{pmatrix}, ~~\alpha = \alpha_0 + \alpha_1 |H_0(\xi)|^2, \notag \\ \notag \\
\mathcal{B}(\xi) &= \begin{pmatrix}
-\textrm{sin}\alpha  & \textrm{cos}\alpha \\
-\textrm{cos}\alpha  & -\textrm{sin}\alpha  \end{pmatrix} \begin{pmatrix}
\textrm{Re}^2H_0  & 0 \\
0 & \textrm{Im}^2H_0  \end{pmatrix}
\notag \\ & ~~~~ +\textrm{Im}H_0\textrm{Re}H_0\begin{pmatrix}
\textrm{cos}\alpha  & -\textrm{sin}\alpha \\
-\textrm{sin}\alpha  & -\textrm{cos}\alpha  \end{pmatrix}, \notag \\ \notag \\
(\mathcal{G}v_i)(\xi)&= \int_{0}^{L}G(\xi-\xi')v_i(\xi')d\xi' ,~~i=1,2 \notag
\end{flalign} where $\textrm{Re}^2H_0 = (\textrm{Re}H_0)^2$ and $\textrm{Im}^2H_0 = (\textrm{Im}H_0)^2$ are also discretized with the same method~\cite{pikovsky_2017}.

\begin{figure}[t!]
\includegraphics[width=1.0\linewidth]{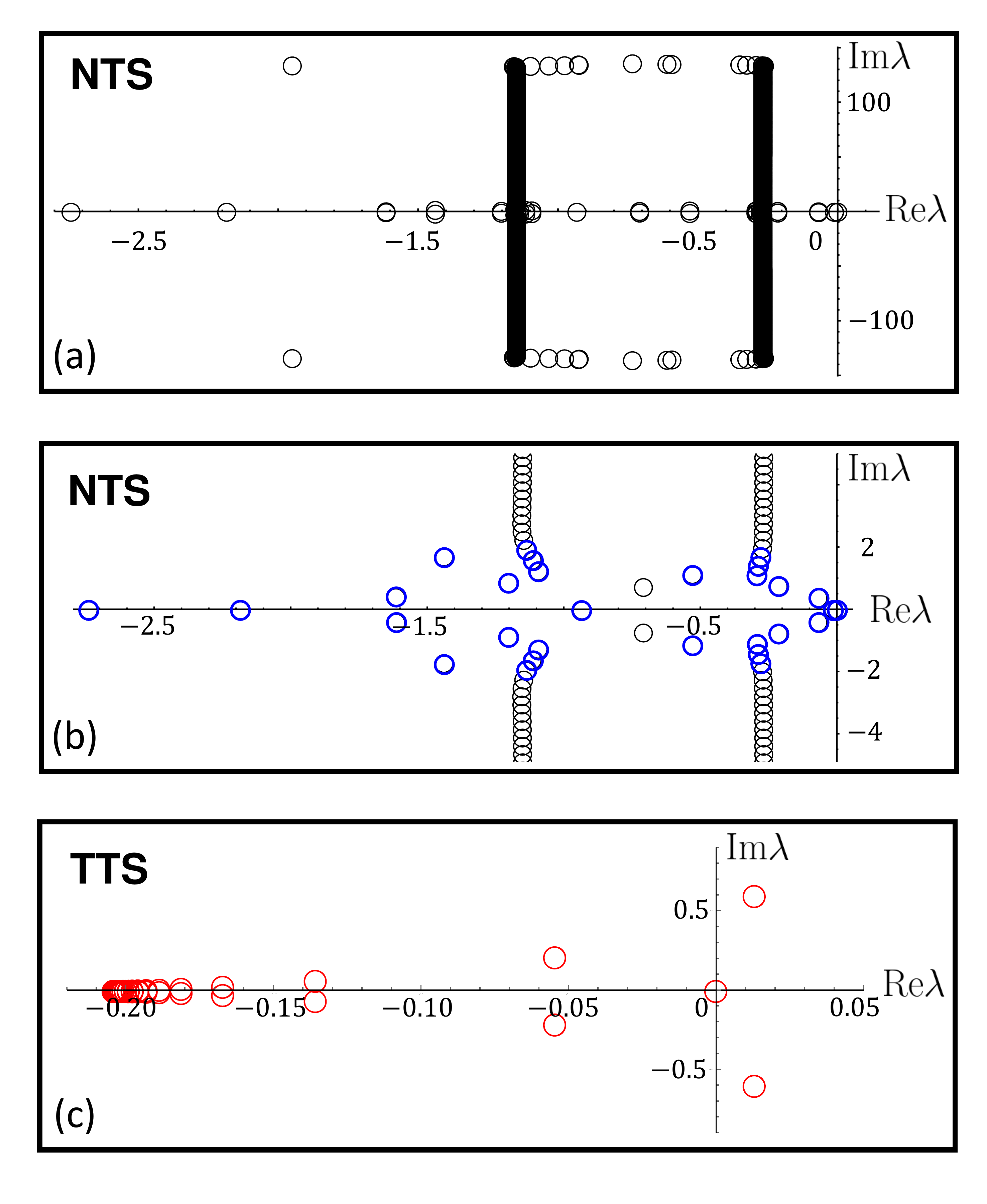}
\caption{\label{Fig:OA-result} (a) Eigenvalues of the linearized system around the NTS solution in a reference frame moving with $c$ and rotating with $\Omega$. (b) Magnification of (a) to highlight the eigenvalues with small imaginary part. The blue dots indicate the eigenvalues which do not coincide with the eigenvalues of the multiplication operator, i.e. $\sigma(\mathcal{L}) \setminus \sigma(\mathcal{M})$, and the black dots display a discretization of the continuous spectrum. These eigenvalues are obtained from the discretization of the linear operator with $M=2^{10}$. (c) Some eigenvalues of the unstable, coexisting TTS near the origin of the complex plane. The parameters are specified in Sec.~\ref{subsec:govern-eqn}.} 
\end{figure}

In Fig.~\ref{Fig:OA-result} (a), the eigenvalues of the linear operator $\mathcal{L}$ are shown in the complex plane. Figure.~\ref{Fig:OA-result} (b) shows a magnification of (a) highlighting the eigenvalues with small imaginary part. The eigenvalues are composed of two branches corresponding to phase and amplitude dynamics, respectively, which can be identified with the discretized form of the continuous spectrum. These continuous branches numerically obtained from the entire linear operator, i.e., $\sigma(\mathcal{L})$ coincide with the eigenvalues of the multiplication operator, i.e., $\sigma(\mathcal{M})$, which means the continuous branches are invariant under the operator $\mathcal{K}$~\cite{Omel_chenko_2013,Omel_chenko_2018}. In addition, there are a few scattered eigenvalues in $\sigma(\mathcal{L})$, which are marked in blue.  According to $\sigma(\mathcal{L}) \setminus \sigma(\mathcal{M})$, we can identify them as  the point spectrum~\cite{Omel_chenko_2013,Omel_chenko_2018}. The latter determines the stability of the NTS. The point spectrum has one zero eigenvalue which comes from the translational invariance and does not affect the stability of the solution. All other eigenvalues have a negative real part, so that the observed NTS is linearly stable. 

At the same parameter values, also a TTS exists. However, its linearization has two complex conjugate eigenvalues with positive real part. It is thus an unstable solution (Fig.~\ref{Fig:OA-result} (c)).

\section{\label{sec:lyapunov}Lyapunov Exponents and Collective Modes}

In a finite-size system, an NTS cannot be represented as a stationary solution in an appropriate reference frame. Rather, we have to treat an NTS as a time-evolving reference trajectory in phase space. Then, we can obtain its spectral properties from a Lyapunov analysis, which yield information about its stability. Therefore, we consider the Jacobian matrix evaluated along a reference trajectory in phase space
\begin{equation}
    (\mathbf{J})_{ij} = \begin{pmatrix}
    \frac{\partial \dot{\phi}_i }{\partial \phi_j} &\vline & \frac{\partial \dot{\phi}_i}{\partial r_j}   \\ \cline{1-3}
    \frac{\partial \dot{r}_i}{\partial \phi_j}&\vline & \frac{\partial \dot{r}_i}{\partial r_j}
    \end{pmatrix} \in \mathbb{R}^{2N\times 2N}, ~~  i,j=1,...,N.
\end{equation} 
Defining the tangent linear propagator $\mathbf{M}(t,t_0)=\mathbf{O}(t)\mathbf{O}^{-1}(t_0)$ where $\mathbf{O}(t)$ is the fundamental matrix solution of $\dot{\mathbf{O}}(t)=\mathbf{J}(t) \mathbf{O}(t)$ with the identity matrix $\mathbf{O}(0) = I_{2N}$~\cite{sjlee1,clv1,clv2}, we obtain the Lyapunov exponents $\Lambda_i$ as an exponential growth rate
\begin{equation}
    \Lambda_i = \lim_{t \rightarrow \infty}\frac{1}{t} \textrm{log}\frac{|| \mathbf{M}(t,t_0) \mathbf{u}(t_0) ||}{||\mathbf{u}(t_0)||}
    \label{Eq:LE-definition}
\end{equation}
along the perturbation vector in the tangent space $\mathbf{T}_{\mathbf{x}^{\textrm{NTS}}(t)}(\mathbb{R}^{2N}) $ where $\mathbf{x}^{\textrm{NTS}}(t)$ is a given NTS reference trajectory in phase space, and $\mathbf{u}(t_0)$ is a perturbation vector belonging to each Oseledets' splitting for $i=1,...,2N$~\cite{clv3,oseledets,LE1}.

\begin{figure}[t!]
\includegraphics[width=1.0\linewidth]{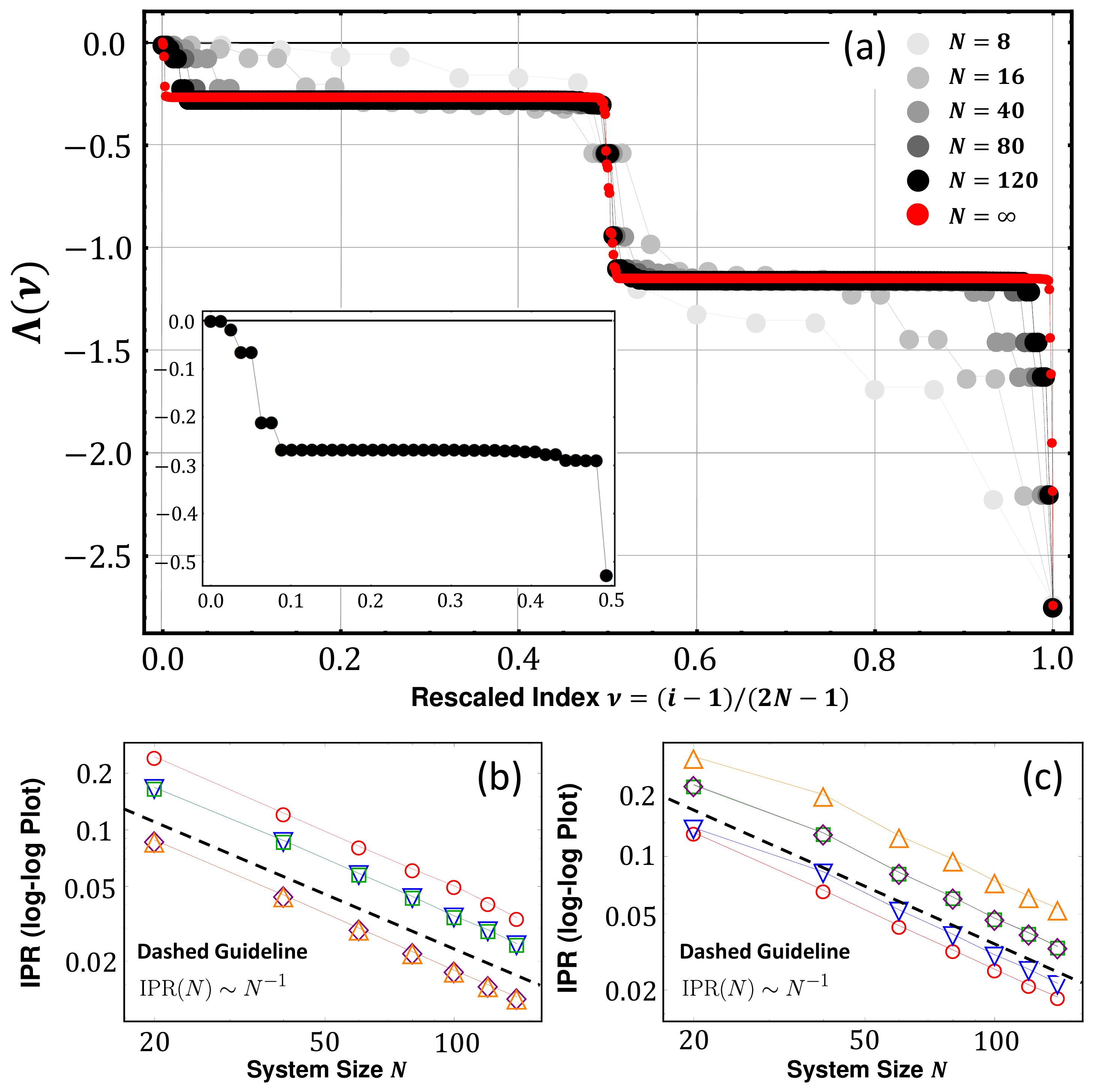}
\caption{\label{Fig:LE-result} 
(a) Lyapunov exponents of the NTS as a function of the re-scaled index $\nu$ for various system sizes. The red dots reproduce the real parts of the eigenvalues obtained from the continuum limit analysis. The inset depicts the first half of the Lyapunov spectrum for $N=40$. (b) and (c) The IPR as a function of system size $N$ corresponding to the discrete Lyapunov exponents around $\nu=0.0$ and $\nu=1.0$, respectively.
} 
\end{figure}

In Fig.~\ref{Fig:LE-result} (a), we show  numerically obtained Lyapunov spectra $\Lambda(\nu)$ for different system sizes $N$ as a function of the re-scaled index $\nu=\frac{i-1}{2N-1}$ (black and gray tone) together with the real part of the point and continuous eigenvalues from the continuum limit analysis (red points). All Lyapunov spectra have two zero Lyapunov exponents which arise from the two continuous symmetries: the time shift invariance due to the autonomous governing equations, and the phase shift invariance, $W \rightarrow W e^{i \chi}$ for $\chi \in \mathbb{R}$, due to the Kuramoto-type phase coupling~\cite{pikovsky_LE,clv3,sjlee1}. These two perturbations do not affect the stability of the NTS reference trajectory. Apart from these two, all other Lyapunov exponents are negative, confirming that the NTS trajectory is stable in all perturbation directions in tangent space. Furthermore, we can distinguish two groups of Lyapunov exponents: discrete Lyapunov exponents and continuous branches, respectively. As apparent from Fig.~\ref{Fig:LE-result} (a), with increasing system size $N$, some of the exponents approach continuous lines with nearly identical values, while others remain discrete points. Among the discrete Lyapunov exponents, some are just single exponents, others form pairs of two nearly-identical Lyapunov exponents, similar to the point spectrum in Sec.~\ref{subsec:stability} which consists of real eigenvalues and pairs of complex conjugate ones, respectively. Taken together, the Lyapunov analysis  strongly suggests that the continuous and discrete parts of the spectrum correspond to the real part of the point and continuous spectra of the continuum limit analysis in Fig.~\ref{Fig:OA-result} (a-b), implying that for $N \rightarrow \infty$ the Lyapunov spectrum converges to the real part of the eigenvalues of the linearized continuum limit equation, as was also found for chimera states ~\cite{chimera-spectral}.

In general, one intuitively expects any propagating wave to be dominated by collective modes since all elements behave in the same way and their entirety forms a propagating structure.
Lyapunov analysis also allows one to measure the `collectivity' of the different Lyapunov modes. The covariant Lyapunov vectors (CLV), which are the spanning set of the Oseledets' splittings, directly indicate the perturbation directions in which the Lyapunov exponents exhibit an exponential growth rate in phase space~\cite{clv1,clv2,oseledets}. From the CLVs, we can derive the time-averaged \textit{inverse participation ratios} (IPRs) for various system sizes, which in turn make it possible to identify collective modes~\cite{collectivemode,clv3,sjlee1}:
\begin{equation}
    \textrm{IPR}^{(i)}(N) = \Bigg< \textrm{exp} \Bigg( \frac{1}{q-1} \textrm{log} \sum_{j=1}^{2N} \bigg| v^{(i)}_{j}(t) \bigg|^{2q} \Bigg) \Bigg>_t
    \label{Eq:ipr-definition}
\end{equation} where $q=2$ and $\textrm{IPR}^{(i)} \in [(2N)^{-1},1]$ and $v_{j}^{(i)}$ is the $j^{\textrm{th}}$ component of the CLV $\mathbf{v}^{(i)}  \in \mathbf{T}_{\mathbf{x}^{\textrm{NTS}}(t)}(\mathbb{R}^{2N}) $ corresponding to a certain Lyapunov exponent $\Lambda_i(N)$ in Eq.~(\ref{Eq:LE-definition}) for $i=1,...,2N$. When the components of a CLV spread out through all the oscillators, $\textrm{IPR}^{(i)}(N) \sim \frac{1}{N}$ as $N \rightarrow \infty$ and the CLV is a collective Lyapunov mode~\cite{collectivemode}. In contrast, when $\textrm{IPR}^{(i)}(N) \sim \textrm{const.}$ as $N$ increases, the vector is well localized.

In Fig.~\ref{Fig:LE-result} (b-c), we show the IPRs as a function of $N$ for the first and the last few Lyapunov modes, which correspond to the discrete LEs near $\nu \gtrsim 0.0$ and $\nu \lesssim 1.0$, respectively. Their IPR decresases as $N$ increases with $\textrm{IPR}(N) \sim \frac{1}{N}$, classifying these modes as collective Lyapunov modes. Also the discrete LEs around $\nu = \frac{i-1}{2N-1} \approx 0.5$ show the same scaling, so that they too are collective Lyapunov modes. From this, we conclude that an NTS trajectory is indeed governed by collective modes that can be captured by  Lyapunov analysis.

\section{\label{sec:hetero}Heterogeneous Natural Frequencies}

\begin{figure}[t!]
\includegraphics[width=1.0\linewidth]{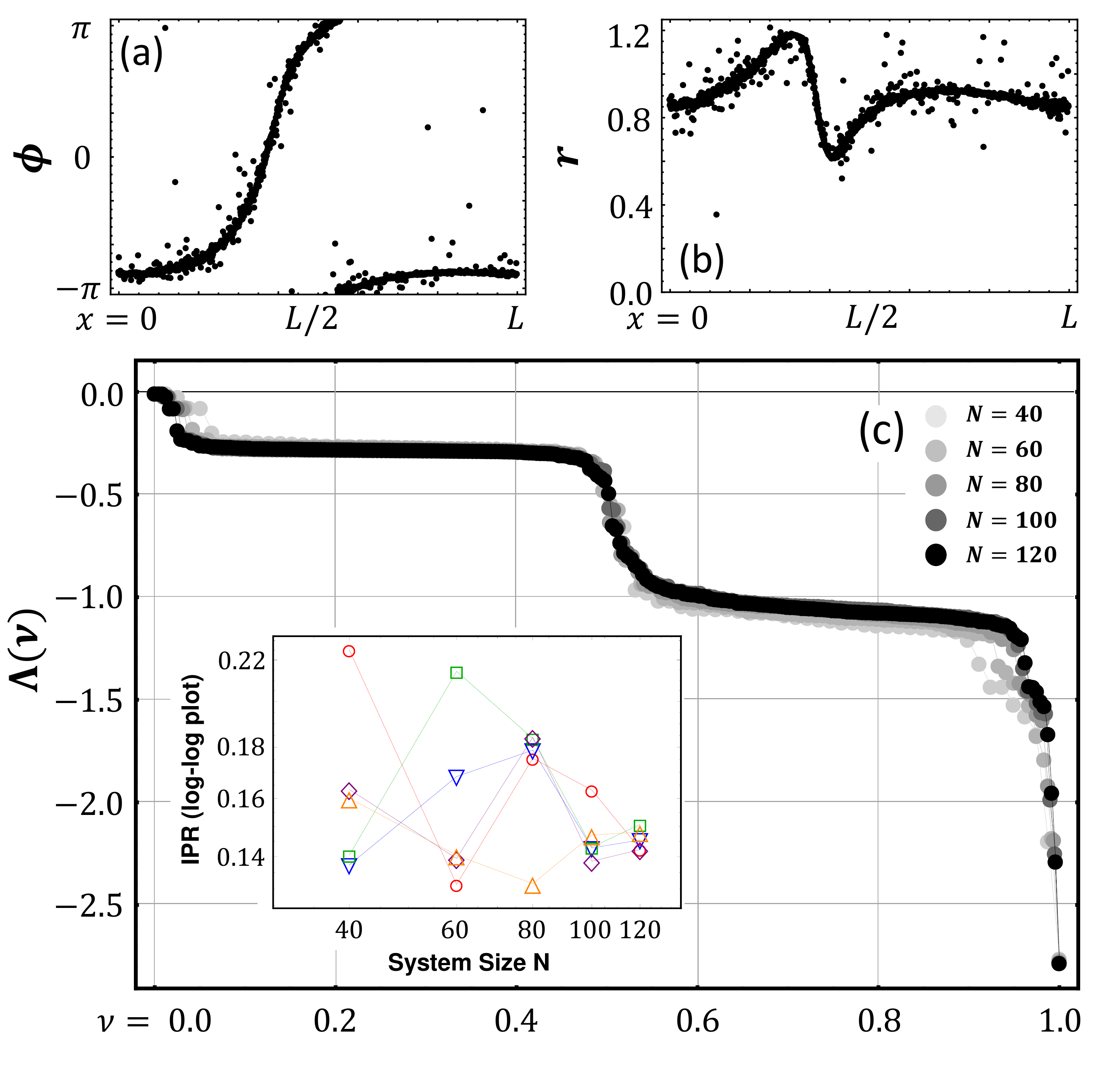}
\caption{\label{Fig:hetero-result} NTS in a system with heterogeneous natural frequencies ($\gamma = 0.002$). (a) Phase snapshot and (b) amplitude snapshot with $N=1600$. (c) Lyapunov exponents for $N=40,60,80,100$, and $120$. Inset: IPR as a function of the system size for the first five modes in $\nu \geq 0.5$.}
\end{figure}

In this section, we demonstrate the robustness of the NTS by adding a small heterogeneity to the natural frequencies of the up to now identical Stuart-Landau oscillators. Therefore, we consider the Cauchy-Lorentz distribution
\begin{equation}
    g(\omega) = \frac{\gamma}{\pi} \frac{1}{\omega^2+\gamma^2} \notag 
\end{equation}
and generate the frequencies according to
\begin{flalign}
\frac{j-\frac{1}{2}}{N} &=\int_{-\infty}^{\tilde{\omega}_{j}}  g(\omega) d\omega = \frac{1}{2}+\frac{1}{\pi}\textrm{tan}^{-1} \big( \frac{\tilde{\omega}_j}{\gamma}\big)
\label{Eq:hetero-distribution}
\end{flalign}
for $j=1,...,N$. 

Then, we mix $\{\tilde{\omega}_j = \gamma \textrm{tan}\big( \frac{\pi(2j-1-N)}{2N} \big) \}_{j=1}^{N}$ to assign one of the randomly distributed natural frequencies to each oscillator $\{\omega_j\}_{j=1}^{N}$ so that $\sum_j \omega_j =0$ exactly. Here, we use $\gamma = 0.002$ to reflect a sufficiently small heterogeneity. 

Fig.~\ref{Fig:hetero-result} (a),(b) depicts numerically obtained snapshots of phase and amplitude profiles of a solution of Eqs.~(\ref{Eq:phase-governing-SLE}-\ref{Eq:amp-governing-SLE}) with heterogeneously distributed natural frequencies. The overall dynamics closely resembles the NTS shown in Fig.~\ref{Fig:TW-snapshot}. However, the profiles are no smooth curves anymore; rather the oscillators are distributed around the wave profile and resemble a partially coherent twisted state reported in Refs.~\cite{omelchenko-twisted,pikovsky-chimera2021,pikovsky-twisted}. Note that this partially coherent NTS only exists for sufficiently small $\gamma$. For example, with $\gamma = 0.02$, we could not find an NTS solution anymore. 

From the Lyapunov spectrum shown in Fig.~\ref{Fig:hetero-result} (c) we can deduce that the NTS remains a stable solution also in the heterogeneous system. All Lyapunov exponents are negative as in the identical system, except for the two zero exponents which also correspond to the two continuous symmetries not influencing the stability of the trajectory. However, the continuous parts of the Lyapunov spectrum seem to differ from the identical case: They are no longer a set of nearly identical values; rather, the values tend to decrease as a function of the re-scaled index. This is reflected in the standard deviation of the LEs between $\nu = 0.6$ and $0.9$, which is approximately $10^{-4}$ (considered zero within our numerical accuracy) for the system of identical oscillators, and approximately $0.028$  for the heterogeneous system. Furthermore, we cannot observe collective modes based on the Lyapunov analysis, as apparent from the inset in Fig.~\ref{Fig:hetero-result} (c): There is no Lyapunov mode whose IPR decreases with $1/N$. The incoherent motion at the microscopic level caused by the heterogeneity of the natural frequencies apparently overshadows the collective response of the oscillators that causes the propagation of the profiles.

\section{\label{sec:bifurcation}Bifurcation Scenario}

\subsection{\label{subsec:one-hump bifurcation} Trivial twisted states and nontrivial twisted state with winding number $q=1$}

In this section, we perform a bifurcation analysis of the TTS and the NTS, based on pseudo-arclength continuation combined with the Newton-Raphson method. The algorithm is described in detail in Ref. \cite{Laing-continuation} and applications can be found in Refs.~\cite{laing-travel1,laing-travel2}. 

\begin{figure}[t!]
\includegraphics[width=1.0\linewidth]{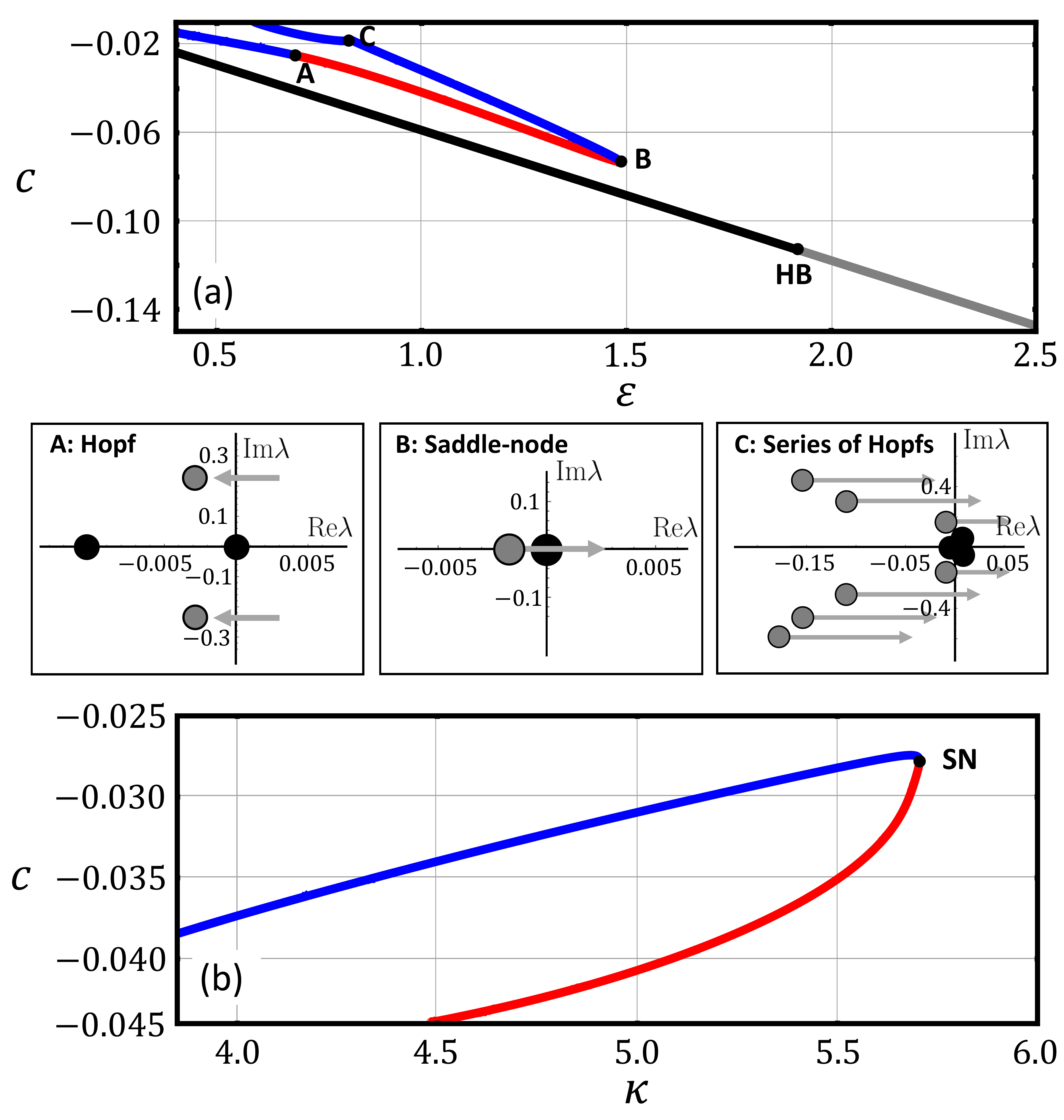}
\caption{ (a) Bifurcation diagrams of the TTS (black and gray lines; black: unstable states, gray: stable states) and NTS (blue and red lines; blue: unstable states, red: stable states) with  $\varepsilon$ the speed $c$ characterizing the NTS solution. The panels in the middle row depict some of the eigenvalues of the linearized equation around the wave profiles in the complex plane close to points (A-C) in (a). (b) Bifurcation diagram of the NTS solution with $\kappa$ as bifurcation parameter. The color code is as in (a). The remaining parameters are the same as in Fig.~\ref{Fig:TW-snapshot}.} 
\label{Fig:bifurcation}
\end{figure}

\begin{figure*}[t!]
\includegraphics[width=1.0\textwidth]{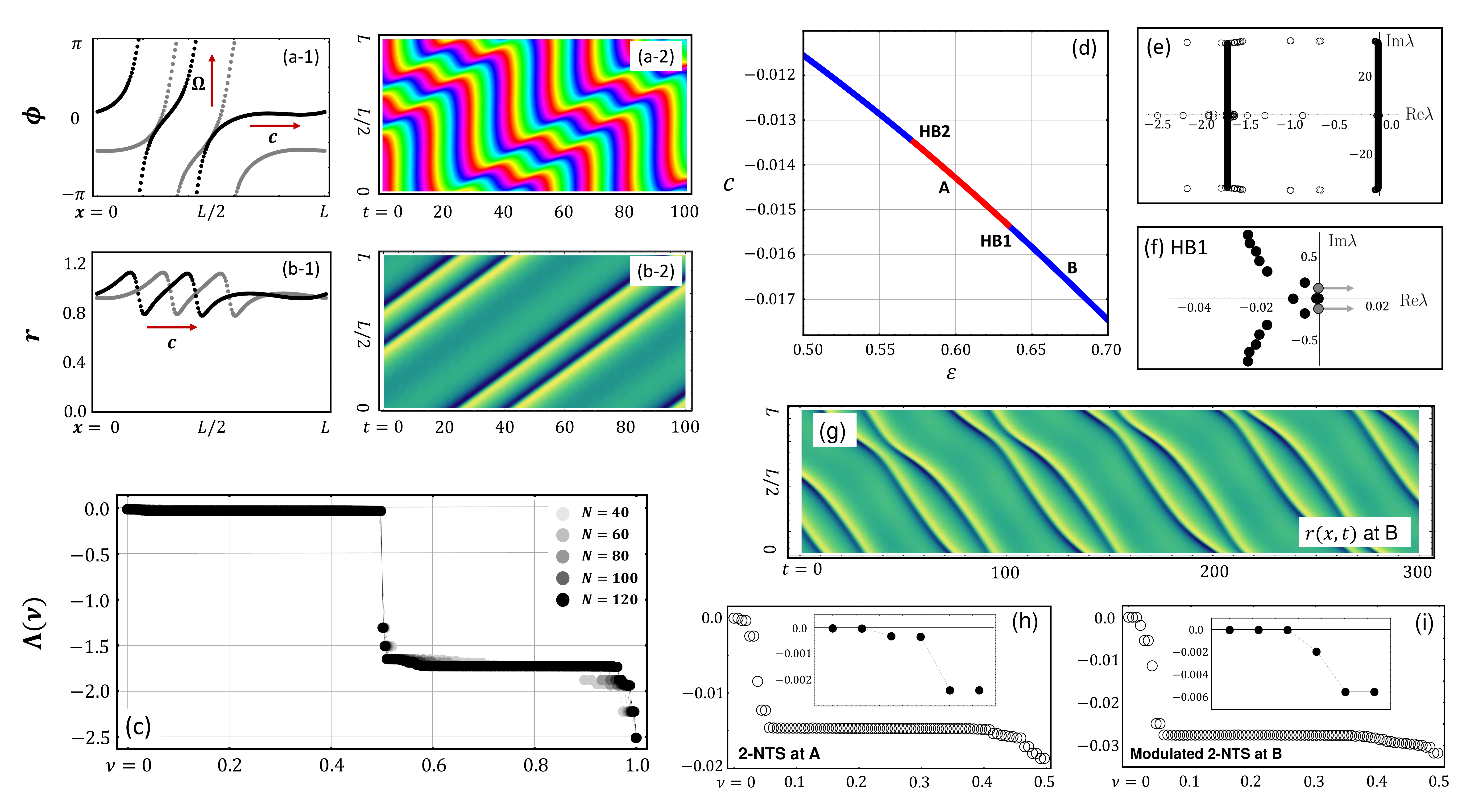}
\caption{ (a-b) Phase and amplitude dynamics of the NTS with $|q|=2$ for $N=200$, and $\varepsilon=0.6$. The other parameters and the color code is the same as in Fig.~\ref{Fig:TW-snapshot}. (c) Lyapunov exponents of the 2-NTS for $N=40,60,80,100$ and $120$. (d) Bifurcation diagrams of the 2-NTS with $\varepsilon$ as bifurcation parameter and the propagation speed $c$ characterizing the NTS. (e) Full eigenvalue spectrum in the complex pane of the Jacobian matrix evaluated at the stationary 2-NTS at \textbf{A}. (f) The eigenvalues near zero in the complex plane at the Hopf bifurcation (\textbf{HB1}). (g) Temporal evolution of the amplitude profile of the modulated 2-NTS at \textbf{B} after \textbf{HB1}; $\varepsilon=0.67$. (h-i) The first half of the Lyapunov exponents of 2-NTS and modulated 2-NTS for $N=80$, respectively. Insets show a magnification close to the origin and highlight the first six Lyapunov exponents.} 
\label{Fig:two-hump}
\end{figure*}

First, we look at a continuation of the TTS with the coupling strength $\varepsilon$ as a bifurcation parameter. In Fig.~\ref{Fig:bifurcation} (a) the TTS is depicted by the black and gray lines, where black indicates unstable TTSs and gray stable ones. The TTS is unstable for low values of $\varepsilon$ and becomes stabilized in a Hopf bifurcation at the point \textbf{HB}, i.e. at a comparatively high value of $\varepsilon$. Our numerical results strongly suggest that the Hopf bifurcation is subcritical. Furthermore, the bifurcation analysis predicts that the velocity $c$ of the TTS depends linearly on $\varepsilon$. This can be easily understood  from the properties of the TTS together with Eq.~(\ref{Eq:phase-governing-SLE}): Since all amplitudes $r_j \approx 1$ and all phase differences of adjacent oscillators are the same, the coupling term in Eq.~(\ref{Eq:phase-governing-SLE}) is identical for all oscillators for a given $\varepsilon$ and scales linearly with $\varepsilon$, resulting in a linear increase of $|c|$ with $\varepsilon$. The branch of unstable TTS continues actually up to $\varepsilon=0$, where also $c=0$.

Let us now focus on the continuation of the NTS with winding number $q=\pm1$ as a function of $\varepsilon$. The corresponding bifurcation diagram is depicted by the red and blue lines in Fig.~\ref{Fig:bifurcation} (a). Stable states are shown in red, unstable ones in blue. Coming from small values of $\varepsilon$, the unstable NTS is stabilized in a Hopf bifurcation at point \textbf{A}, as judged from the course of the eigenvalues with $\varepsilon$ in the complex plane (see the left panel in the middle row of Fig.~\ref{Fig:bifurcation}).
Beyond the Hopf bifurcation, the stable NTS exists in a large $\varepsilon$ interval, ensuring that its existence is not restricted to a practically inaccessible parameter range. Along this stable curve, the difference between the maximum and minimum values of the amplitude hump increases with increasing $\varepsilon$. The fully synchronized state is stable for $\varepsilon \gtrapprox  0.773$. Thus, the NTS coexists with the uniform oscillation in most of its existence range. However, in 30 simulations with random initial conditions, 28 and 29 trajectories approached the NTS state at $\varepsilon=1.0$ and $\varepsilon=1.2$, respectively. This suggests that the basin of attraction of the NTS is considerably larger than the one of the synchronized oscillation.

At the high-$\varepsilon$ end of its existence interval, the NTS solution is annihilated in a saddle-node bifurcation (SN) at point \textbf{B} (cf. the middle panel in the middle row).  Continuing the unstable NTS branch that is born in the SN bifurcation \textbf{B}, we observe that it is further destabilized in a series of Hopf bifurcations starting at \textbf{C}. At the same time, the amplitude profile becomes flatter and flatter approaching the uniform profile as $c$ approaches zero.

Obviously, stable NTS and stable TTS do not coexist. Their existence range is separated by the $\varepsilon$ interval between points \textbf{B} and \textbf{HB} where none of them is stable. In this interval, more precisely, in 100 numerical integration with $\varepsilon=1.6$ as well as $\varepsilon=1.8$ and random initial conditions, the trajectory always approached the fully synchronized oscillation. In comparison, when doing the same numerical experiment for $\varepsilon=2.0$ and $\varepsilon=2.2$, $95 \%$ of the initial conditions end up on the TTS and only $5 \%$ on the uniform oscillation. For $\varepsilon=1.0$ and $\varepsilon=1.2$, where the NTS is stable, none and 2 out of 30 simulations with random initial conditions approach the uniform state, respectively. Furthermore, in the shown range of $\varepsilon$ the bifurcation diagrams for TTS and NTS remain well separated, suggesting that the two solutions do not interact directly at any bifurcation point. We note that we were not able to continue the NTS for values of $\varepsilon$ smaller than the ones shown in Fig.~\ref{Fig:bifurcation} due to convergence problems.

Besides the region between \textbf{B} and \textbf{HB}, there is a second region in which neither NTS nor TTS are stable, namely for $\varepsilon < \varepsilon(\textbf{A})$. In this parameter interval we observed states with discontinuous amplitude dynamics. Thus, here, the amplitudes do not form a smooth curve. Examples include states where the phase dynamics seems similar to the one of the irregular inhomogeneous states reported in Ref.~\cite{pikovsky-chimera2018}, amplitude-mediated chimera states~\cite{amc}, as well as different kinds of NTS solutions, as discussed in the next subsection.

Finally, a continuation of the NTS with the parameter $\kappa$ (the inverse of the interaction range) reveals that the solution is restricted to a certain interaction range, or certain length of the system (Fig.~\ref{Fig:bifurcation} (b)).

\subsection{\label{subsec:two-hump bifurcation} Nontrivial twisted states with winding number $q=2$}

So far, we only discussed an NTS with a winding number $|q|=1$. However, as apparent from the definition, just as a TTS, an NTS may also have a winding number $|q|>1$. In this section, we address an NTS solution of Eq.~(\ref{Eq:complex-governing}) with winding number $|q|=2$, and call it a 2-NTS. In Fig.~\ref{Fig:two-hump} (a-b), such a 2-NTS solution is depicted. The amplitude profile features two humps and the phase profile winds twice along the ring, changing in total by $4\pi$. From the Lyapunov exponents, we can conjecture that the 2-NTS, too, is a stable solution. However, it is less stable than the 1-NTS since the first half of the Lyapunov exponents are closer to zero than in the case of the 1-NTS (compare Fig.~\ref{Fig:two-hump} (c),(h) to Fig.~\ref{Fig:LE-result} (a)). This can be also verified by determining the eigenvalues of the linearized equation in the continuum limit analysis (Fig.~\ref{Fig:two-hump} (e)): One of the continuous parts of the spectrum of the 2-NTS is closer to the imaginary axis than that of the 1-NTS.

A bifurcation analysis reveals that the stable 2-NTS solution can be observed in some parameter region, ensuring it is a robust solution (Fig.~\ref{Fig:two-hump} (d)). However, the parameter interval is smaller than the one in which the stable 1-NTS solution exists. Moreover, it is found at lower values of the coupling strength $\varepsilon \approx 0.6$, and its speed tends to be lower than the one of the stable 1-NTS. At both ends of its existence interval the stable 2-NTS becomes destabilized  through Hopf bifurcations (cf. Fig.~\ref{Fig:two-hump} (f) for HB1). Furthermore, we numerically verified that one of them (\textbf{HB1}) is indeed a supercritical Hopf bifurcation. Before \textbf{HB1}, the 2-NTS solution exhibits a stationary amplitude profile in a moving reference frame (Fig.~\ref{Fig:two-hump} (b-2)). Beyond \textbf{HB1}, e.g. at $\varepsilon=0.67$, a 2-NTS state is still observed, but now its amplitude profile oscillates periodically, rendering the state a modulated traveling wave (Fig.~\ref{Fig:two-hump} (g)). The modulated NTS features three zero Lyapunov exponents, two of which arise from the continuous symmetries, the third originating from the modulation frequency (Fig.~\ref{Fig:two-hump} (i)).

\section{\label{sec:conclusion}Discussion and Conclusion}

In this work, we reported a new type of collective behavior in a ring of nonlocally coupled Stuart-Landau oscillators, which we named a nontrivial twisted state. It is characterized by non-uniform profiles of amplitude and phase gradient as well as a winding number $q$. The latter characterizes it as a twisted state and implies that the structure is a coherent traveling wave. From a macroscopic point of view, the modulus of the local order parameter of an NTS (see Eq.~(\ref{Eq: local order paramter})) $|z(x,t)|=1$ for all $x \in [0,L]$ and all $t$ while the global order parameter (see Eq.~(\ref{Eq:global-Kuramoto-OP})) $0 < |\Gamma(t)| <1$ for all $t$. In contrast, the well-known trivial twisted state has a constant phase gradient and uniform amplitude profile, which renders the global order parameter zero $|\Gamma(t)|=0$ while the modulus of the local order parameter remains 1, $|z(x,t)|=1$ for all $x$ and $t$. Linear stability analysis, Lyapunov analysis and bifurcation analysis revealed that NTS solutions with winding number $|q| = 1,2$ are attracting states which exist in wide ranges of parameter sets and for many initial conditions.

In the literature, there are some examples of coherent spatio-temporal patterns of coupled oscillators in a ring geometry that resemble an NTS in some respect. Most of them were observed in studies of phase models. A coherent traveling wave solution in a model of coupled phase oscillators with a non-constant phase gradient profile was reported in \cite{co-travel} (see Fig.~24). However, it is not a twisted state since the difference of phases does not add up to an integer multiple of $2\pi$ but rather to $0$. In other words, the phase does not form complete cycles when going once around the ring so that one cannot assign a winding number. In the Kuramoto-Sakaguchi phase model with nonlinear phase-lag function, initial conditions close to a TTS led transiently to an evolution of the phase profile that resembles the one in our NTS solutions before settling down to a chimera state. However, this NTS-like dynamics was not obtained as a stable state (see Fig.~2 (c) in \cite{pikovsky-twisted}). In the Kuramoto-Sakaguchi phase-reduced model with a trigonometric nonlocal coupling kernel, coherent wave solutions with non-constant phase gradient profile coming closest to the ones discussed here are reported (Fig.~1 (c) in \cite{Omel_chenko_2018}). However, these traveling waves are again not stable solutions but long-lived transients with many neutrally stable directions~\footnote{O. E. Omel'chenko, Personal Communication, May 19, 2022}. Finally, in a model of Stuart-Landau oscillators with time-delay, a traveling wave solution possessing a winding number and a slightly varying phase gradient was found (Fig.~2 (d) in \cite{scholl-travel2}), so that the modulus of the global order parameter was close to zero, yet finite.  

As we indicated in Sec.~\ref{sec:bifurcation}, besides the NTS solutions, the system of nonlocally coupled Stuart-Landau oscillators on a ring seems to possess various kinds of further coherent traveling wave solutions. Many of them represent apparently novel types of collective behaviors such as a solitary state presenting discontinuities at some locations in an otherwise smooth profile.  
In the future, further studies of nonlocally coupled Stuart-Landau oscillators may therefore reveal other novel types of collective behaviors. 

Finally, it is worthwhile also to compare the NTS to the partial synchrony observed in globally coupled oscillators~\cite{partial1,partial2,partial3, nakagawa1,nakagawa2} as a splay state in the globally coupled system is similar to the TTS in the spatially extended system. The amplitude profiles of some partially synchronized states in globally coupled oscillators form a smooth closed curve as a function of phase $\phi \in [-\pi,\pi)$ in the complex plane~\cite{politi-quasi}. Similarly, the amplitude of the NTS forms a smooth closed curve in the complex plane as a function of spatial variable $x \in [0,L]$. In both cases, the individual oscillators behave quasi-periodically while the collective dynamics is periodic. However, a prominent difference between them seems to be that the partial synchrony in globally coupled systems bifurcates from the splay state whereas the NTS does not bifurcate from the TTS but rather emerges in a saddle-node bifurcation.

\begin{acknowledgments}
The authors would like to thank O. E. Omel'chenko and C. R. Laing for fruitful discussions. This work has been supported by the Deutsche Forschungsgemeinschaft (project KR1189/18 ‘Chimera States and Beyond’)
\end{acknowledgments}

\bibliography{apssamp}

\end{document}